\documentstyle[preprint,aps]{revtex}                                 

\newcommand{\be}{\begin{equation}}
\newcommand{\ee}{\end{equation}}
\newcommand{\bea}{\begin{eqnarray}}
\newcommand{\eea}{\end{eqnarray}}

\begin{document}                                                     
\draft                                                               
\title{                                                              
Partially Solvable Anisotropic {\it t-J} Model with Long-Range Interactions    
}                                                                    
\author{Yasuhiro {\sc Saiga}, Yusuke {\sc Kato} and Yoshio {\sc Kuramoto}}                        
\address{Department of Physics, Tohoku University,\\                 
Sendai 980-77, Japan}                                                
\maketitle
                                                           
\begin{abstract}
A new anisotropic $t$-$J$ model in one dimension is proposed which has long-range hopping and exchange. 
This $t$-$J$ model is only partially solvable in contrast to known integrable models with long-range interaction. 
In the high-density limit the model reduces to the XXZ chain with the long-range exchange.
Some exact eigenfunctions are shown to be of Jastrow-type if certain conditions for an anisotropy parameter are satisfied.  
The ground state as well as the excitation spectrum for various cases of the anisotropy parameter and filling are derived numerically.
It is found that the Jastrow-type wave function is an excellent trial function for any value of the anisotropy parameter.
\end{abstract}


\newpage

There have been remarkable developments in understanding the family of one-dimensional models with interaction proportional to the inverse square of the distance.\cite{Sutherland71,Sutherland72,Haldane,Shastry,K-Y,WLC,K-K}
The simple structure of the spectrum places the family as canonical models for realizing the Tomonaga-Luttinger liquid in one dimension.
Much less is known for the case where an energy gap is present for the excitation spectrum.  The simplest example is the XXZ chain with the Ising anisotropy.  
It has been shown by Haldane \cite{Haldane} that for particular values of the anisotropy parameter, the long-range XXZ chain has the Jastrow-type wave function as an eigenfunction with non-zero magnetization.
However, the wave function of the ground state is {\it not} of Jastrow type.  The presence of the gap makes the system different from the Tomonaga-Luttinger liquid.
In view of importance of the anisotropy in real physical systems, the study on the anisotropic models is urgently required.

The purpose of this Letter is to present and solve a new type of the {\it t-J} model which reduces to the long-range XXZ chain in the high-density limit.
We can derive a part of the eigenstates by mapping the model to the multicomponent Sutherland model.~\cite{K-K}  
The ground-state eigenfunction within a certain range of the magnetization is shown to be of Jastrow type.  We clarify the condition under which the Jastrow-type function constitutes an eigenstate.
For cases where analytic solution is not available, we diagonalize the Hamiltonian numerically.  The ground-state wave function so obtained turns out to be extremely close to but not identical with the Jastrow-type one.
As a preliminary to understanding of the full spectrum, we examine the excitation spectrum of the long-range XXZ model numerically.


Let us first review the anisotropic spin model (XXZ model) with long-range exchange.~\cite{Haldane}  
The Hamiltonian is given by 
\bea
  {\cal H} &=& \sum_{i \ne j}^N J_{ij} \bigl( S_i^x S_j^x + S_i^y S_j^y + \Delta S_i^z S_j^z \bigr), \label{eq:Hamiltonian} 
\eea
with $J_{ij} = JD(x_i-x_j)^{-2}$, where $D(x_i-x_j) = (N/\pi) \left| \sin [\pi (x_i-x_j)/N] \right|$ is the chord distance consistent with the periodic boundary condition on $N$ lattice sites. 
The parameter $\Delta(\ge 1)$ represents the Ising anisotropy. 
Haldane~\cite{Haldane} has shown that the model is partially soluble under some conditions on anisotropy and magnetization, and has obtained a Jastrow-type eigenfunction and its eigenenergy. 
The model we present below is an extension of the XXZ model with inclusion of holes.


We introduce an anisotropic $t$-$J$ model as follows:
\begin{eqnarray}
  {\cal H} &=& {\cal P} \sum_{i \neq j} \biggl\{ - t_{ij} \left( c_{i \uparrow}^\dagger c_{j \uparrow} + \Delta_1 c_{i \downarrow}^\dagger c_{j \downarrow} \right) + J_{ij} \biggl[ S_i^x S_j^x + S_i^y S_j^y + \Delta_2 S_i^z S_j^z \nonumber \\
           & & + \left( \Delta_3 - 1 \right) \left( S_i^z n_j + n_i S_j^z \right) - \frac{1}{4} \Delta_4 n_i  n_j \biggr] \biggr\} {\cal P}, \label{Hamiltonian}
\end{eqnarray}
where ${\cal P}$ is the projection operator to exclude the double occupation at each site. 
We take $t_{ij}=J_{ij}=tD(x_i-x_j)^{-2}$, where $D(x_i-x_j)$ is defined above.
The parameters $\Delta_1$ and $\Delta_2$ obviously break the SU(2) invariance.  The other parameters $\Delta_3$ and $\Delta_4$ are necessary to make the model analytically solvable.
It may seem highly unrealistic that the model has different magnitudes for the hopping.  This difference originates simply from the mathematical requirement.  If we translate the spin degrees of freedom into the chain index in a double-chain model, however, the difference is physically acceptable.
When $\Delta_i=1$ for all $i$, this model reduces to the long-range supersymmetric $t$-$J$ model.~\cite{K-Y}  


Any state in the Hilbert space can be represented by the positions of holes and down spins.
If we let $M$ denote the number of down spins and $Q$ denote the number of holes, then $S_{\rm tot}^z$ is given by $S_{\rm tot}^z=(N-Q)/2-M$. 
The wave function is represented by
\begin{eqnarray}
  |\psi \rangle &=& \sum_{\{ x \},\{ y \}} \psi(\{ x \};\{ y \}) \prod_{i = 1}^M S_{x_i}^- \prod_{j=1}^Q h_{y_j}^\dagger |F\rangle, 
\end{eqnarray}
where $|F\rangle$ denotes the fully polarized up-spin state, 
$S_{x_i}^-$ is the spin-lowering operator at site $x_i$, and $h_{y_j}^\dagger$ creates a hole at site $y_j$.
The amplitude $\psi(\{ x \};\{ y \})$ is symmetric in the positions $\{x\}=\{x_1, \cdots, x_M\}$ of down spins, and antisymmetric in the positions $\{y\}=\{y_1, \cdots, y_Q\}$ of the holes. 
When $\{x\}$ and $\{y\}$ are specified, the positions $\{u\}=\{u_1, \cdots, u_{N-M-Q}\}$ of the up spins are determined automatically. 

To solve the eigenvalue problem ${\cal H}|\psi\rangle=E|\psi\rangle$, 
we divide (\ref{Hamiltonian}) as follows: 
$ {\cal H} = {\cal H'} + {\cal H}_{\rm rest}$, 
where 
\begin{eqnarray}
  {\cal H}_{\rm rest} &=& {\cal P} \sum_{i \neq j}  \biggl[ \frac{1}{2} J_{ij} ( \Delta_2 + 2 \Delta_3 -2 ) \left( S_i^z + S_j^z \right) 
- \frac{1}{4} J_{ij} ( \Delta_2 + 2 \Delta_3 - 2 ) \nonumber \\
                      & & - \frac{1}{4} J_{ij} ( \Delta_4 - 2 \Delta_3 + 2 ) ( n_i + n_j )
+ \frac{1}{4} J_{ij} ( \Delta_4 - 2 \Delta_3 + 2 ) \biggr] {\cal P}. 
\end{eqnarray}
This part ${\cal H}_{\rm rest}$ is diagonal i.e., ${\cal H}_{\rm rest}|\psi\rangle=E_{\rm rest}|\psi\rangle$, and gives a constant term $E_{\rm rest}$ for the eigenvalue.  Namely we obtain
\begin{eqnarray}
  E_{\rm rest} &=& \frac{\pi^2 t}{12 N^2} (N^2 - 1) \left[ (\Delta_2 + 2 \Delta_3 - 2) (N - 4M -2Q) - (\Delta_4 - 2 \Delta_3 + 2) (N - 2Q) \right]. 
\end{eqnarray}
Next, we consider ${\cal H'}$ in the Hilbert space with the reference state $|F\rangle$.  It is represented by 
\begin{eqnarray}
  {\cal H'} &=& - \sum_{i \in \{ y \}} \sum_{j \in \{ u \}} t_{ij} h_i h_j^\dagger - \Delta_1 \sum_{i \in \{ y \}} \sum_{j \in \{ x \}} t_{ij} S_i^- h_i h_j^\dagger S_j^+ \nonumber \\
& & + \sum_{i \in \{ x \}} \sum_{j \in \{ u \}} J_{ij} S_i^+ S_j^- + (\Delta_2 + 2 \Delta_3 - 2) \sum_{i \notin \{ u \}} \sum_{j \notin \{ u \},j \neq i} J_{ij} \left( S_i^z - \frac{1}{2} \right) \left( S_j^z - \frac{1}{2} \right) \nonumber \\
& & - 2 ( \Delta_3 - 1 ) \sum_{i \in \{ x \}} \sum_{j \in \{ x \},j \neq i} J_{ij} n_i n_j \left( S_i^z - \frac{1}{2} \right) \left( S_i^z - \frac{1}{2} \right) \nonumber \\
& & - \frac{1}{4} ( \Delta_4 - 2 \Delta_3 + 2 ) \sum_{i \in \{ y \}} \sum_{j \in \{ y \},j \neq i} J_{ij} (n_i - 1)(n_j - 1). 
\end{eqnarray}
In order to make the model soluble we parametrize $\Delta_i$'s as $\Delta_1 = \lambda$, $\Delta_2 = \lambda (\lambda + 1)/2$, $\Delta_3 - 1 = \lambda (\lambda - 1)/4$, and $\Delta_4 = \lambda (3 - \lambda)/2$ in terms of a single parameter $\lambda$. Then ${\cal H}|\psi\rangle=E|\psi\rangle$ turns into the following equation: 
\be
  \sum_{i=1}^{M+Q} \sum_{r=1}^{N-1} J_r \psi(v_1,\cdots,v_i + r,\cdots,v_{M+Q}) + \sum_{i<j} \lambda (\lambda + P_{ij}) J_{ij} \psi(\{ v \}) = (E - E_{\rm rest}) \psi(\{ v \}). \label{tjeigeneq1}
\ee
where we use the notation $\{ v \} = \{ x \} \oplus \{ y \}$, and
\begin{equation}  
  E_{\rm rest} = \frac{\pi^2 t}{6 N^2} (N^2 - 1) [\lambda (\lambda - 1) (N - 2Q) - 2 \lambda^2 M].
\end{equation}

With the plain wave basis, a general wave function can be represented as 
\be
  \psi(\{ v \}) = \prod_{i=1}^{M+Q} z_i^{N/2} \sum_{\{ k \}} z_1^{k_1} \cdots z_{M+Q}^{k_{M+Q}} \psi_{\rm F} (\{ k \}) \equiv \prod_{i=1}^{M+Q} z_i^{N/2} \cdot \tilde{\psi}, \label{wavefunction}
\ee
where F of $\psi_{\rm F}$ means the Fourier expansion and $z_i \equiv \exp({\rm i} 2 \pi v_i/N)$. 
We shall specify the range of the momenta $k_i (i=1, \cdots, M+Q)$ of two kinds of particles (down spins and holes) shortly.
Substituting (\ref{wavefunction}) into (\ref{tjeigeneq1}), we obtain
\begin{eqnarray}
  2 \left[ \sum_{i=1}^{M+Q} \left( z_i \frac{\partial}{\partial z_i} \right)^2 -2 \sum_{i<j} \frac{ z_i z_j \lambda (\lambda + P_{ij}) }{(z_i - z_j)^2} \right] \tilde{\psi} = \left[ \frac{N^2}{\pi^2 t} \left( E - E_{\rm rest} \right) + \frac{N^2 + 2}{6} (M + Q) \right] \tilde{\psi}. \label{tjeigeneq2} 
\end{eqnarray}
Here we have used the relation:
\be
  \sum_{r=1}^{N-1} \frac{e^{{\rm i} 2 \pi k r/N} }{\sin^2 (\pi r/N)} = 2 \left( k - \frac{N}{2} \right)^2 - \frac{N^2 + 2}{6}
\ee
for $0 \le k \le N$. 
In order to derive (\ref{tjeigeneq2}), it is thus required that $-N/2 \le k_i  \le N/2$. 
Since the effect of umklapp scattering is not considered in $\tilde{\psi}$, we are not dealing with the complete set of wave functions. 
The left-hand side of (\ref{tjeigeneq2}) is nothing but the Hamiltonian of the multicomponent Sutherland model with both fermions and bosons present.~\cite{K-K}  
Thus the results of the multicomponent Sutherland model can be used to find the solution of the anisotropic $t$-$J$ model. 


A Jastrow-type wave function 
\be
  \psi(\{ x \};\{ y \}) = \prod_{i=1}^M e^{{\rm i} 2 \pi x_i J_{\rm s}/N} \prod_{j=1}^Q e^{{\rm i} 2 \pi y_j J_{\rm h}/N} \prod_{i<j}^M D(x_i - x_j)^{\lambda + 1} \prod_{i<j}^Q D(y_i - y_j)^\lambda \prod_{i=1}^M \prod_{j=1}^Q D(x_i - y_j)^\lambda \label{anisotjwavefunc}
\ee
is an eigenfunction of (\ref{Hamiltonian}) where $\Delta_i$'s are parametrized with $\lambda$, under the following conditions: $\lambda$=odd integer, $\left| J_{\rm s} - J_{\rm h} \right| \le (M+1)/2$, and 
\bea
  & & \left| J_{\rm s} - \frac{N}{2} \right| \le \frac{N}{2} - \frac{1}{2} \lambda(M+Q-1) - \frac{1}{2} (M-1), \\
  & & \left| J_{\rm h} - \frac{N}{2} \right| \le \frac{N}{2} - \frac{1}{2} \lambda(M+Q-1).
  \eea
Here $J_{\rm s}$ and $J_{\rm h}$ represent the uniform currents of down spins and holes, respectively. 
They are related with total momentum $P_{\rm tot}=(2 \pi/N)(J_{\rm s} M + J_{\rm h} Q)$. 
The condition $\lambda$=(odd integer) comes from the statistics of particles: down spins are hard-core bosons and holes are fermions.
The three inequalities given above stem from the restrictions in mapping to the multicomponent Sutherland model. 
If the above conditions are satisfied, the wave function (\ref{anisotjwavefunc}) yields the following eigenenergy: 
\bea
  \frac{N^2}{\pi^2 t} E &=& \frac{1}{6} (\lambda + 1)^2 M (M^2 - 1) - 2 M J_{\rm s} (N - J_{\rm s})
                        + \frac{1}{3} (N^2 - 1) \left[ \frac{\lambda (\lambda - 1)}{2} N + (1 - \lambda^2) M \right] \nonumber \\
                        & & + \frac{1}{6} (\lambda + 1)^2 Q (Q^2 - 1) - 2 Q J_{\rm h} (N - J_{\rm h})
                        + 2 \lambda Q (J_{\rm s} - J_{\rm h})^2 - \frac{1}{3} (\lambda^2 - \lambda - 1) (N^2 - 1) Q \nonumber \\
                        & & + \frac{1}{2} Q \biggl\{ \lambda (M + Q) [ (\lambda + 1) M - \lambda Q ]
                        + (\lambda - 1) \left[ \lambda M Q + \left( \lambda + \frac{1}{3} \right) Q^2 - \frac{1}{3} \right] \biggr\}. \label{anisotjenergy} 
\eea
When $\lambda=1$ i.e., $\Delta_i=1$ for all $i$, this reduces to the result by Wang {\it et al}.~\cite{WLC}  
We obtain the ground state with magnetization by choosing both $J_{\rm s}$ and $J_{\rm h}$ as close as possible to $N/2$, while respecting the restrictions on $M$, $Q$, $J_{\rm s}$, and $J_{\rm h}$.

Mapping (\ref{Hamiltonian}) to the multicomponent Sutherland model allows us to derive some excited states in addition to the states given here. 
Owing to the restrictions in the mapping, however, not all of states are found analytically in the anisotropic $t$-$J$ model. 
The same applies to the long-range XXZ model. 
Thus the anisotropic models are different from completely soluble models: Bethe-ansatz solvable models or the $1/\sin^2$-type isotropic models. 
We emphasize that the anisotropic models belong to a new family of partially soluble models. 

                                                                      
We remark on the absolute ground state when the magnetization is      
varied.                                                               
When $\lambda > 1$, the state cannot be found analytically. 
 Numerical calculation shows that the state has   
$S_{\rm tot}^z \ne 0$ with the parameter $\lambda$ larger than $\lambda_{\rm c} = 2.25 \pm 0.05$ for $(N, Q)=(8, 2)$.   
This is because the term with $(\Delta_3 - 1)$ in (\ref{Hamiltonian}) 
breaks the time reversal symmetry and acts like a magnetic field.     
In contrast to the spin model, presence of carriers in the {\it t-J}  
model should always lead to the gapless spectrum.                     
We have also studied the case where the term with $(\Delta_3 - 1)$ in 
(\ref{Hamiltonian}) is omitted.   The absolute ground state has then  
$S_{\rm tot}^z =0$ up to $\lambda=10$ for $(N, Q)=(8, 2)$ and $(8, 6)$ 
according to the numerical diagonalization. We remark that the        
ground-state eigenfunction is no longer of Jastrow-type without the   
$(\Delta_3 - 1)$ term.                                                


We have seen that the Jastrow-type wave functions are exact solutions of the anisotropic models provided that the anisotropy parameters satisfy the solvable conditions. 
We now calculate the overlap 
$|\langle \psi_{\rm exact} |\psi_{\rm J}\rangle|$
between the exact eigenfunction $\psi_{\rm exact}$ and the Jastrow-type wave function $\psi_{\rm J}$ in order to see the validity of the latter function as a trial function beyond the condition for the exact solution.

Figure \ref{overlapspin} shows the overlap against the Ising anisotropy parameter $\Delta$ for 12 sites of the spin model. 
We first note that the overlap is almost unity over the whole range of the anisotropy. 
For example, its value is 0.97693 for $\Delta=10$ and $S_{\rm tot}^z=0$. 
This means that the Jastrow-type wave function constitutes an excellent approximation even if the Ising anisotropy does not satisfy the condition for the analytic solution. 
In addition, the values of $\Delta$ where the overlap is unity are: (a)$\Delta=1$; (b)$\Delta=$1 and 6. 
These $\Delta$ satisfy the condition for the analytic solution. 
Second, comparing the top figure (a) with the bottom one (b), we find that the overlap in the case of $S_{\rm tot}^z =2$ is closer to unity than that for $S_{\rm tot}^z=0$. 
Finally, the overlap as a function of $\Delta$ is nonmonotonic with a conspicuous minimum and a faint second minimum.
This behavior for $S_{\rm tot}^z=0$ is common to other cases with different number of sites such as $N=6, 8$ and $10$.

One may naturally ask whether the Jastrow-type function remains a good trial function if holes are introduced. 
Figure \ref{overlaptj} shows the overlap against the anisotropy parameter $\lambda$  for the $t$-$J$ model with 8 sites. 
Note that the anisotropy parameter is not $\Delta_i$'s but $\lambda$. 
It is evident that the overlap is also very close to unity for the whole range of $\lambda$. 
For $\lambda=10, M=1$ and $Q=2$, for instance, its value is 0.99965. 
In the case where $\lambda=$1 and 3 with $(N, M, Q)=(8, 1, 2)$, the solvable conditions are satisfied. 
Hence the overlaps are unity in that case. 
In the limit of large $\lambda$ the {\it t-J} model describes localized holes and down spins.  Then the ground state is a regular lattice of three components: up and down spins, and holes.  
As in the case of the spin model, the overlap shows a nonmonotonic dependence on the anisotropy parameter.

One of the reasons why the Jastrow-type function is so good a trial function is the following:  As $\lambda$ increases the wave function described by (\ref{anisotjwavefunc}) becomes sharply peaked for evenly spaced coordinates of down spins and holes. 
The center of gravity is moving with a definite value of the total momentum. This state becomes the exact ground state of the Hamiltonian in the limit of large $\lambda$.  Here the ground-state energy does not depend on the total momentum, as seen by the absence of $J_{\rm h}$ and $J_{\rm s}$ terms of the leading order $O(\lambda^2)$ in (\ref{anisotjenergy}).  
Thus the Jastrow-type function (\ref{anisotjwavefunc}) represents the exact eigenstate not only in the isotropic limit, but also in the limit of large $\lambda$. 
It is still surprising that the Jastrow-type function interpolates the two limits so precisely. 


We now discuss the excitation spectrum of the present anisotropic model.
One may ask whether the concept of spinons and holons remains useful in the case of anisotropic models.  As the first step toward detailed investigation of the spectrum of the {\it t-J} model, we use numerical diagonalization to derive the excitation energy of the XXZ chain for each value of the total momentum $K$.
Figure \ref{excitationspin} shows the spectrum for the system with 12 sites and $\Delta=10$. 
It is evident that there is a quasi-continuum of excitations with energy gaps from the ground state.
This continuum has a strong resemblance to the result of the nearest-neighbor spin model by Ishimura and Shiba~\cite{IS}.  The almost vanishing excitation gap ($= 3.3008 \times 10^{-3} J$) 
at $K= \pi$ reflects the degenerate ground state in the Ising limit.

In summary, we have constructed the anisotropic $t$-$J$ model which is analytically solvable for a restricted set of states.
The model is probably {\it not} a completely integrable model unlike the other isotropic long-range models. 
We have shown that the Jastrow-type wave function is appropriate as the trial function over the whole range of the anisotropy parameter and filling.
The excitation spectrum of the $t$-$J$ model will be discussed elsewhere.

\begin{figure}[r]                                                     
\caption{The overlap vs the Ising anisotropy $\Delta$ for the anisotropic spin model. 
The total azimuthal spin $S_{\rm tot}^z=N/2-M$ is: (a)$S_{\rm tot}^z=0$ and (b)$S_{\rm tot}^z=2$. 
$N$ and $M$ denote the numbers of sites and down spins, respectively. 
The minimum and the maximum are shown by up and down triangle, respectively. } 
\label{overlapspin}                                                    
\vspace{3cm}                                                          
\end{figure}                                                          

\begin{figure}[r]
\caption{The overlap vs the anisotropy parameter $\lambda$ for the anisotropic $t$-$J$ model. 
The total azimuthal spin $S_{\rm tot}^z=(N-Q)/2-M$ is 2. 
$N$, $M$ and $Q$ denote the numbers of sites, down spins and holes, respectively. }
\label{overlaptj}
\vspace{3cm}
\end{figure}

\begin{figure}[r]
\caption{The excitation spectrum of the anisotropic spin chain model with $N=12$, $\Delta=10$ and $S_{\rm tot}^z=0$. $E_{\rm g}$ represents the ground-state energy. }
\label{excitationspin}
\end{figure}

\end{document}